\documentclass[twocolumn,prb,showpacs,preprintnumbers,amsmath,amssymb,floatfix]{revtex4-1}
\usepackage{graphicx}
\usepackage{amsmath}
\usepackage{amssymb}
\usepackage{color}

\begin{document}

\title{Theory of percolation and tunneling regimes in nanogranular metal films}

\author{Claudio Grimaldi}
\affiliation{Laboratory of Physics of Complex Matter, 
Ecole Polytechnique F\'ed\'erale de Lausanne, Station 3, CP-1015 Lausanne, Switzerland}

\begin{abstract}
Nanogranular metal composites, consisting of immiscible metallic and insulating phases deposited on a substrate, are characterized
by two distinct electronic transport regimes depending on the relative amount of the metallic phase. At sufficiently large
metallic loadings, granular metals behave as percolating systems with a well-defined critical concentration
above which macroscopic clusters of physically connected conductive particles span the entire sample.
Below the critical loading, granular metal films are in the dielectric regime, where current can flow throughout 
the composite only via hopping or tunneling processes between isolated nanosized particles or clusters. 
In this case transport is intrinsically non-percolative in the sense that no critical concentration can be identified
for the onset of transport.
It is shown here that, although being very different in nature, these two regimes can be described by
treating percolation and hopping on equal footing. By considering general features of the microstructure and of the 
electrical connectedness, the concentration dependence of the dc conductivity of several 
nanogranular metal films is reproduced to high accuracy within an effective medium approach. In particular, fits
to published experimental data enable us to extract the values of microscopic parameters that govern the percolation
and tunneling regimes, explaining thus the transport properties observed in nanogranular metal films.

\end{abstract}


\maketitle

\section{Introduction}
\label{intro}

Among the different classes of conductor-insulator composites, nanogranular metal films
are quite unique materials as they display distinct and tunable electrical, optical, and 
magnetic properties depending on the nature and concentration of the metallic phase, as well
as on the structure of the films.\cite{Au-Al2O3,Niklasson1984,Batlle2002} In the preparation of this class of 
composites, various sputtering, evaporation, and ion implantation methods are used 
to deposit immiscible metals and insulators on a 
substrate to form composite films with a wide range of the volume fraction $\phi$ of the metallic phase.  
At large $\phi$,
the composite is basically a metallic continuum whose electronic conductivity $\sigma$ is limited 
by grain boundaries and scattering with few insulating inclusions. 
As $\phi$ decreases, $\sigma$ is lowered by the enhanced concentration of the insulating phase. 
This ``metallic regime'' persists until matrix inversion occurs at a material dependent
critical value $\phi_c$, below which the metallic continuum is broken up into disconnected
metallic particles or clusters dispersed in the insulating phase. In this ``dielectric regime'', electrons flow
throughout the composite only by tunneling or hopping processes between isolated, and homogeneously
dispersed, nanometric metallic particles.

These two distinct, structurally driven, transport regimes are very different in nature, as seen from
the temperature and relative concentration dependences of $\sigma$. Above $\phi_c$,
granular metal films are assimilable to percolative systems in which coalescing metallic particles
form a system spanning conductive network. As a function of $\phi$, the resulting conductivity 
for $\phi\gtrsim\phi_c$ is thus expected to follow a percolation power-law behavior of the 
form:\cite{Stauffer1994,Sahimi2003}
\begin{equation}
\label{power}
\sigma \simeq \sigma_0 (\phi-\phi_c)^t,
\end{equation}
where $\sigma_0$ is a constant and $t\simeq 2$ ($t\simeq 1.3$) is the universal transport exponent for
three-dimensional (two-dimensional) systems. Furthermore, for $\phi>\phi_c$ transport
shows typically a metallic behavior, with the resistivity increasing linearly with the temperature.
In contrast to the percolation mechanism implied by Eq.~\eqref{power}, tunneling between submicron
conducting particles in the dielectric regime hints to 
the absence of any ``critical'' concentration, as electrons have to tunnel across interparticle 
distances that increase gradually as $\phi$ is reduced.
Considering that in the limit of dilute particles of size $D$ the mean particle 
separation $\delta$ scales as $\delta\propto D/\phi^{1/d}$, where $d$ is the system dimensionality,
the tunneling conductivity $\sigma\propto\exp(-2\delta/\xi)$ for sufficiently large temperatures 
is thus expected to follow:
\begin{equation}
\label{expo}
\sigma\propto\exp\!\left(-a_d\frac{D}{\xi\phi^{1/d}}\right),
\end{equation}
where $\xi$ is the tunneling decay length and $a_d$ is a dimensionless constant, which for point particles 
dispersed in a three-dimensional (two-dimensional) volume takes up the value $a_3\simeq 1.41$ 
($a_2\simeq 2.12$).\cite{Hunt2005,Seager1974} As a function of temperature $T$, the dielectric regime 
is associated with a stretched exponential behavior of the form:
\begin{equation}
\label{sigmaT}
\sigma\propto\exp\!\left(-\sqrt{\frac{T_0}{T}}\right),
\end{equation}
which arises from tunneling processes in the presence of a Coulomb gap.\cite{Efros1975,Shklovskii1988,Beloborodov2007,Pollak2013}
Equation~\eqref{sigmaT} applies for temperatures lower than a $\phi$-dependent characteristic temperature, $T_0$, which typically
increases from $T_0\sim 100$ K for $\phi\lesssim\phi_c$ to a few thousands of Kelvin for $\phi$
values deep in the dielectric region.\cite{Au-Al2O3,Fe-SiO2}

The limiting  $\phi$-dependencies of $\sigma$ highlighted in Eqs.~\eqref{power} and \eqref{expo} arise from 
general considerations which do not rely on the detailed knowledge of the composite film morphology.
However, while Eq.~\eqref{power} is shown to properly fit the measured
$\sigma$ in the metallic regime,\cite{McAlister1985,Ni-SiO2,Ag-SnO2,Ag-Al2O3,Salvadori2013} 
the exponential behavior of Eq.~\eqref{expo} 
is less often used to interpret the observed $\phi$-dependence of the dielectric regime.\cite{Ag-SnO2,Fostner2014} 
In Ref.~\onlinecite{Ni-SiO2}, for example, the dielectric region is understood in terms of Eq.~\eqref{power} with
a tunneling-induced nonuniversal exponent,\cite{Balberg1987} while in Refs.~\onlinecite{Ag-SnO2,Ag-Al2O3} 
additional percolation transitions are considered to be active in the $\phi<\phi_c$ region.
The power law behavior of Eq.~\eqref{power}, or its generalizations, requires however either a cut-off in the inter-particle
conductances,\cite{Ambrosetti2010a} or very peculiar (crystal-like) arrangements of the metallic particles in 
the matrix,\cite{Ambrosetti2010b} both of which are difficult to justify from the disordered morphology of
nanogranular films in the dielectric regime. Furthermore, the variable range hopping mechanism at the basis
of the stretched exponential behavior of Eq.~\eqref{sigmaT} is, in principle, incompatible with the notion of
a fixed cut-off distance between the particles.  
From these considerations, we see that although the temperature dependence of $\sigma$ is quite well
understood,\cite{Beloborodov2007} there is still no general consensus on how to interpret the behavior of the conductivity
as a function of the metallic content at fixed temperatures. In particular, there is a need to
further understand the different regimes of granular metals within a single, coherent, description.

In this paper, we present an effective medium formulation that naturally accounts for the metallic (percolation) and
dielectric (tunneling) regimes of granular thick films, and the transition between them. With the term ``thick film''
we mean that the film thickness is much larger than the typical particle size, so that the system is
three dimensional. By considering general properties
of the microstructure and of the electrical connectedness, we clarify how the percolation behavior of 
Eq.~\eqref{power} for $\phi>\phi_c$  evolves into the exponential one of Eq.~\eqref{expo} for 
$\phi<\phi_c$. In this way, we can reproduce the room temperature conductivity data of several composite films in the whole 
range of $\phi$, and extract from experiments the tunneling characteristics, the percolation threshold, and the 
microscopic conductances governing the overall conductivity behavior.
Furthermore, by using a cherry-pit model for the conductive particle dispersion in the films, we identify the
different observed values of $\phi_c$ in terms of partial overlaps between the particles, providing thus a
simple microscopic interpretation for the location of the dielectric-metallic transition.

\section{Model and Effective medium approximation}
\label{ema}

\begin{figure}[t]
\begin{center}
\includegraphics[scale=1,clip=true]{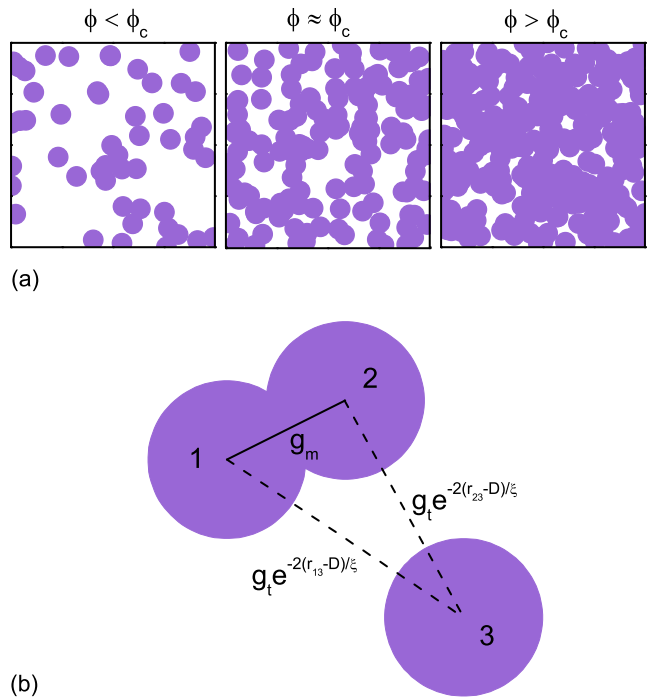}
\caption{(Color online) (a) Illustration of the model for nanogranular composite films
in which the metallic particles are represented as partially overlapping spheres dispersed 
in a continuum matrix. For volume fractions $\phi$ lower than the percolation threshold $\phi_c$, 
the composite is constituted by dispersions of isolated particles and clusters of overlapping
particles. For $\phi>\phi_c$ the film microstructure is modeled as a continuum of overlapping spheres
with isolated voids. $\phi_c$ is the critical volume fraction for the percolation of overlapping spheres.
(b) Model of inter-particle conductances. For any two overlapping spheres, the conductance is set equal
to $g_m$, as for spheres $1$ and $2$. When two particles do not overlap, their conductance is chosen to
be of tunneling type, as for spheres $1$ and $3$, and $2$ and $3$.}
\label{fig1}
\end{center}
\end{figure}
\label{matrix}

During the deposition process of granular metal films, the metallic particles nucleate and grow
giving rise to a spatial distribution of the metallic grains which depends on the relative amount
of metallic and insulating phases, on the interaction between them, and on the film growth conditions. 
Detailed description of film microstructure requires thus specific knowledge of the microscopic processes governing
the spatial distribution of the two phases. However, observational studies of many different film 
microstructures evidence quite general features, such as homogeneity and disorder of particle dispersions,
nanometric size of metallic grains in the dielectric regime, and matrix inversion in the transition
region. These generic features can be taken into account without detailed knowledge of the processes 
governing them by employing a minimal model of the microstructure chosen as to capture the 
essential aspects. 

To define a minimal model for granular metal films we consider spherical metallic
particles with equal diameter $D$ dispersed in a continuum insulating medium. We simulate coalescing between
the metallic particles by allowing the spheres to overlap to some extent. At low volume fractions, the
metallic phase is thus composed primarily of isolated spheres, while at large $\phi$ the composite consists
basically of a metallic continuum with few isolated voids, as shown schematically in Fig.~\ref{fig1}(a).
The regions of low and large $\phi$ correspond respectively to the dielectric and metallic regimes
of the granular films. The critical volume fraction $\phi_c$ separating these two regions corresponds
to the geometrical percolation threshold for intersecting spheres, i.e.,  $\phi_c$ is the smallest $\phi$ such
that a system spanning cluster of overlapping spheres exists. The specific value of $\phi_c$ depends on the 
degree of particle overlapping and on the statistical properties of the dispersion, which however we do not
specify at the moment. 

To model the electrical connectedness at the microscopic level, we define two kinds of interparticle conductances, 
as illustrated schematically in Fig.~\ref{fig1}(b).
When two particles overlap, as for example particles $1$ and $2$ in Fig.~\ref{fig1}(a), we assume that the
interparticle conductance is constant and independent of the degree of overlapping. Between nonoverlapping particles,
instead, we ascribe a tunneling conductance which decays exponentially with the relative distance between the particles,
as for the pairs of spheres $1,3$ and $2,3$ in Fig.~\ref{fig1}(b). For any two given spheres $i$ and $j$, 
the interparticle conductance assumes thus the following form:
\begin{equation}
\label{modelg}
g_{ij}=\left\{
\begin{array}{lll}
g_m & \textrm{for} & r_{ij}\leq D, \\
g_t e^{-2(r_{ij}-D)/\xi} & \textrm{for} & r_{ij}> D,
\end{array}\right.
\end{equation}
where $r_{ij}$ is the distance between the sphere centers and $\xi$ is the
tunneling decay length. 
The prefactors $g_m$ and $g_t$ in Eq.~\eqref{modelg}
are in general different: $g_t$ is in principle smaller or much smaller than the conductance 
of two coalesced particles as the electron has to cross an interfacial barrier even if two particles are
at contact. Furthermore, for the case of nanosized ferromagnetic particles, $g_t$ depends 
also on the relative spin  polarization.\cite{Inoue1996}

In writing Eq.~\eqref{modelg} we assume that particle charging and
Coulomb interaction effects do not appreciably contribute to the exponential decay for $r_{ij}>D$.
Although strictly valid for infinite temperatures, this approximation is nevertheless justified for granular metals
at room temperature and with metallic contents
not much below $\phi_c$, as in this case the variable range hopping characteristic temperature
in Eq.~\eqref{sigmaT} is typically $T_0\sim 100$ K.\cite{Fe-SiO2} Since we are interested in the $\phi$-dependence
of $\sigma$ at room temperature, we can think of the prefactor $g_t$ as to partially include particle 
charging and Coulomb interactions. 

\subsection{Effective medium approximation}
\label{secema}
We proceed to evaluating the composite film conductivity by using an effective medium 
approximation (EMA) previously applied successfully to a number of different conductor-insulator
composites.\cite{Ambrosetti2010b,Grimaldi2011,Nigro2012a,Nigro2012b,Nigro2013b}
A detailed derivation of EMA is presented in Ref.~\onlinecite{Grimaldi2011}. Here we describe
a simple method to derive the EMA equation within the two-site approximation
by considering a system of $N$ metallic spheres dispersed in a volume $V$. We then  
construct a resistor network whose node positions coincide with the centers of the spherical particles.
The corresponding bond conductances are given
by the set of $N(N-1)/2$ conductances of Eq.~\eqref{modelg}. This network is complete, which means that
to each pair of nodes is associated a finite conductance $g_{ij}$. The two-point resistance $R_{ij}$ between
any two nodes $i$ and $j$ is thus a well defined quantity, from which we construct the average
resistance of the network:
\begin{equation}
\label{appa1}
\langle R\rangle=\frac{1}{N(N-1)}\left\langle\sum_{i,j}{}'R_{ij}\right\rangle,
\end{equation}
where $\langle\cdots\rangle$ indicates a configurational average
and the prime symbol means that the term with $i=j$ is omitted from the summation. 
We can express $R_{ij}$ as given by the direct resistance between $i$ and $j$, i.e., $1/g_{ij}$, in parallel with 
the resistance $1/G'_{ij}$ of a network in which $g_{ij}$ has been removed from the system:
\begin{equation}
\label{appa2}
R_{ij}=\frac{1}{g_{ij}+G'_{ij}}.
\end{equation}
Next, we introduce a second (effective) network, with spatial distribution of nodes identical to the original one, in which the
conductances are all identically equal to $\bar{g}$, independently of the node indexes. This second network is a
complete network whose two-point resistance is simply given by $\bar{R}=1/\bar{G}=2/N\bar{g}$.\cite{Wu2004} 
We want to find $\bar{g}$ such that the resistance difference between the two networks
\begin{equation}
\label{appa3}
\langle R\rangle-\bar{R}=\frac{1}{N(N-1)}\left\langle\sum_{i,j}{}'\left(\frac{1}{g_{ij}+G'_{ij}}-\frac{2}{N\bar{g}}\right)\right\rangle
\end{equation}
vanishes. To this end, we apply the two-site EMA which amounts to replacing 
$G'_{ij}$ by the two-point conductance of the effective network minus the direct contribution between $i$ and $j$:
\begin{equation}
\label{appa4}
G'_{ij}\rightarrow  \bar{G}-\bar{g}=(N/2-1)\bar{g},
\end{equation}
so that Eq.~\eqref{appa3} reduces to:
\begin{equation}
\label{appa5}
\frac{\langle R\rangle-\bar{R}}{\bar{R}}=\frac{1}{N(N-1)}\left\langle\sum_{i,j}{}'\frac{\bar{g}-g_{ij}}{g_{ij}+(N/2-1)\bar{g}}\right\rangle.
\end{equation}
Imposing $\langle R\rangle=\bar{R}$ to the above expression, after some algebra and 
setting $N\gg 1$ we find the following equation for the effective conductance $\bar{G}$:\cite{Ambrosetti2010b,Grimaldi2011}
\begin{equation}
\label{appa6}
\frac{1}{N}\left\langle\sum_{i,j}{}'\frac{g_{ij}}{g_{ij}+\bar{G}}\right\rangle=2.
\end{equation}
Since the conductances $g_{ij}$ in Eq.~\eqref{modelg} depend only upon the relative distances $r_{ij}$, 
we can replace the summation over $i,j$ by an integral over the continuous distance $r$.\cite{Ambrosetti2010b,Grimaldi2011} 
By using Eq.~\eqref{modelg} we thus obtain for three dimensional systems:
\begin{equation}
\label{ema1}
Z(\phi,D)\frac{g_m}{\bar{G}+g_m}
+4\pi\rho\!\int_D^\infty\!dr r^2 g_2(r)\frac{g_t e^{-\frac{2(r-D)}{\xi}}}{\bar{G}+g_te^{-\frac{2(r-D)}{\xi}}}=2,
\end{equation}
where 
\begin{equation}
\label{rdf}
g_2(r)=\int\frac{d\Omega}{4\pi}\left\langle\frac{1}{N\rho}\sum_{i,j}{}'\delta(\mathbf{r}-\mathbf{r}_{ij})\right\rangle
\end{equation}
is the radial distribution function for the conducting spheres,\cite{Hansen2006} and
\begin{equation}
\label{zeta}
Z(\phi,D)=4\pi\rho\int_0^D\!dr r^2 g_2(r)
\end{equation}
is the coordination number for intersecting spheres, which measures how many spheres on average overlap
a given sphere for a given concentration.\cite{TorquatoBook} Finally, $\rho=N/V$ is the particle number density which,
depending on the degree of sphere overlapping, determines the fractional coverage $\phi$ of 
the metallic phase.

\subsection{EMA dielectric and metallic regimes}
\label{limiting}
Equations \eqref{ema1} and \eqref{zeta} enable us to relate the behavior of the overall transport
with the morphology of the composite through the radial distribution function $g_2(r)$, once this is known.
Detailed knowledge of $g_2(r)$ is however not necessary to extract some important limiting behaviors
of $\bar{G}$ from the solution of Eq.~\eqref{ema1}.
For example, the EMA dielectric regime is obtained by noticing that in the 
dilute limit $\phi\ll 1$ the metallic particles are uncorrelated [$g_2(r)\simeq 1$] and practically do not 
overlap [$Z(\phi, D)\ll 1$]. In this way, the first term in the left-hand side of Eq.\eqref{ema1} can be neglected and
the EMA equation reduces to:
\begin{equation}
\label{expo2}
\frac{24\phi}{D^3}\int_D^\infty \!dr r^2\frac{g_t e^{-\frac{2(r-D)}{\xi}}}{\bar{G}+g_te^{-\frac{2(r-D)}{\xi}}}=2,
\end{equation}
where we have set $\phi\simeq \pi\rho D^3/6$. The above integral is exactly solvable and the left-hand side
of Eq.~\eqref{expo2}, which we denote by $I$, can be expressed in terms of polylogarithm functions. 
We find it more practical, however, to use for $I$ the following approximation which is very accurate for all values 
of $\bar{G}$: 
\begin{equation}
\label{I}
I=8\phi\left\{\left[1+\frac{\xi}{2D}\ln\!\left(\frac{g_t+\bar{G}}{\bar{G}}\right)\right]^3-1\right\}.
\end{equation}
From $I=2$ we thus find for small $\phi$:
\begin{align}
\label{expo3}
\bar{G}&\simeq g_t\exp\left\{-\frac{2D}{\xi}\left[\left(\frac{1}{4\phi}+1\right)^{1/3}-1\right]\right\}\nonumber \\
&\underset{\phi\rightarrow 0}{\longrightarrow} g_t\exp\left(-1.26\frac{D}{\xi\phi^{1/3}}\right),
\end{align}
which has the same asymptotic behavior of Eq.~\eqref{expo}. Equation~\eqref{expo3}
can also be recovered from the method described in Ref.~\onlinecite{Ambrosetti2010b}.

To obtain the EMA version for the percolating regime, we neglect the tunneling contributions in Eq.~\eqref{ema1}.
The EMA equation reduces to $Z(\phi,D)g_m/(\bar{G}+g_m)=2$, from which we find:
\begin{equation}
\label{power2}
\bar{G}=g_m\left[\frac{Z(\phi,D)}{2}-1\right].
\end{equation}
Since $Z(\phi,D)$ increases monotonically with $\phi$, $\bar{G}$ is non-negative
only for $\phi\geq \phi_c$, where $\phi_c$ satisfies $Z(\phi_c,D)=2$. Hence, by
expanding Eq.~\eqref{power2} in the vicinity of $\phi_c$ we obtain for $\phi-\phi_c\gtrsim 0$:
\begin{equation}
\label{power3}
\bar{G}\simeq \frac{g_m Z'(\phi_c,D)}{2}(\phi-\phi_c),
\end{equation}
which is the EMA equivalent of the percolation conductivity of Eq.~\eqref{power}, in which the
transport exponent is unity rather than $t\simeq 2$.

\section{Minimal EMA model for nanogranular metal films}
\label{minimal}

\begin{figure}[t]
\begin{center}
\includegraphics[scale=0.78,clip=true]{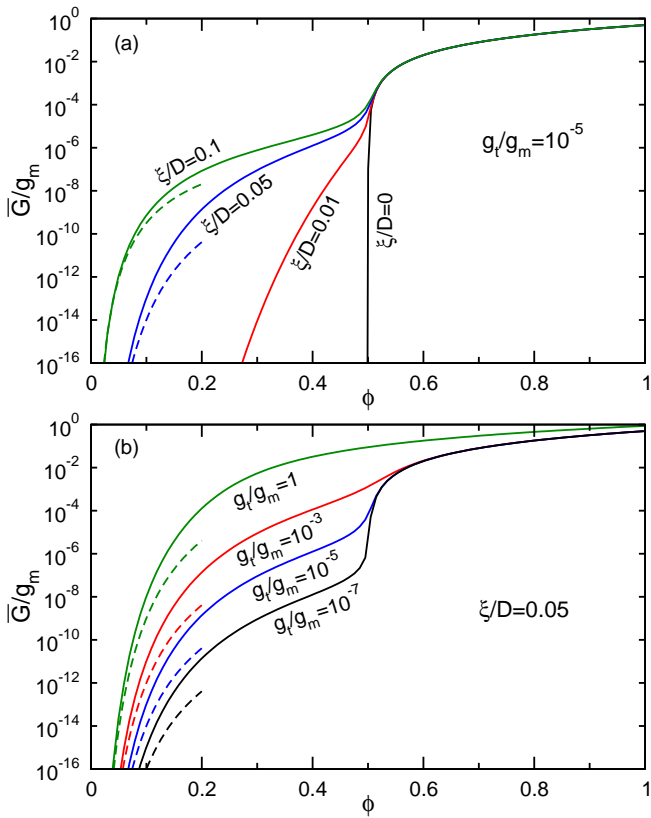}
\caption{(Color online) EMA conductance $\bar{G}$ as a function of the volume fraction $\phi$ of conducting spheres
as obtained from numerical solutions of Eq.~\eqref{emamin}.
The critical volume fraction $\phi_c$ for overlapping spheres is fixed at $\phi_c=0.5$.
(a) $\bar{G}$ for different values of $\xi/D$ at fixed $g_t/g_m=10^{-5}$. For $\xi/D=0$ the EMA conductance follows
the percolation behavior $\bar{G}\propto (\phi-\phi_c)^t$, with $t=2$. Dashed lines are the low density
$\bar{G}$ of Eq.~\eqref{expo3}.
(b) $\bar{G}$ for different values of $g_t/g_m$ at fixed $\xi/D=0.05$. }
\label{fig2}
\end{center}
\end{figure}

From the results of the previous section, we are now in the position of formulating a
minimal, phenomenological model describing the $\phi$ dependence of the conductivity
of nanogranular metal films. The starting point is Eq.~\eqref{ema1}, which we modify in the
following way. First, motivated by the observation that the microstructure of the film is expected to have little 
influence in the dilute particle limit $\phi\ll 1$,  where tunneling dominates, we replace the second term in the left-hand side 
of \eqref{ema1} with Eq.~\eqref{I}.
Next, to keep the number of independent parameters to an
absolute minimum, we assume a simple linear dependence of the coordination number:
$Z(\phi,D)=b\phi$. In this way, the critical volume fraction is uniquely identified by $\phi_c=2/b$,
as can be verified by using Eq.~\eqref{power2}. Finally, to recover the correct exponent
in the percolating regime, we follow the phenomenological approach of Ref.~\onlinecite{McLachlan1989} 
and replace the quantity $g_m/(\bar{G}+g_m)$ in Eq.~\eqref{ema1}
with $g_m^{1/t}/(\bar{G}^{1/t}+g_m^{1/t})$, where we set $t=2$ for three dimensional materials.
The resulting EMA equation reduces thus to:
\begin{equation}
\label{emamin}
\frac{(\phi/\phi_c)g_m^{1/t}}{\bar{G}^{1/t}+g_m^{1/t}}+
4\phi\left\{\left[1+\frac{\xi}{2D}\ln\!\left(\frac{g_t+\bar{G}}{\bar{G}}\right)\right]^3-1\right\}=1.
\end{equation}
It is easy to see from the above equation that for $\phi\ll\phi_c$ the EMA conductance reduces
to Eq.~\eqref{expo3}, while for $\phi \gtrsim\phi_c$ (and for sufficiently small $\xi/D$) it
takes the percolation form $\bar{G}\propto(\phi-\phi_c)^t$. It is worth stressing that
while the tunneling contribution is treated explicitly, the percolation threshold is used as a
parameter of the theory, with no explicit relation with the specific microstructure. In this respect,
compared to the model of semi-penetrable spheres introduced in Sec.~\ref{ema}, Eq.~\eqref{emamin} 
represents a semi-phenomenological description of nanocomposite films.

The $\phi$-dependence of $\bar{G}$, obtained by numerical solution of Eq.~\eqref{emamin}, is
shown in Fig.~\ref{fig2} for different values of $\xi/D$ and $g_t/g_m$, with critical volume
fraction fixed at $\phi_c=1/2$. For $\xi/D=0$, transport is purely percolative and the EMA conductance
follows $\bar{G}=(g_m/\phi_c^t)(\phi-\phi_c)^t$ for $\phi\geq \phi_c$. In this region, the percolating behavior 
persists even for $\xi/D\neq 0$, while for $\phi<\phi_c$ the tunneling contributions become dominant
and $\bar{G}$ asymptotically follows Eq.~\eqref{expo3}, as shown in Fig.~\ref{fig2}(a).
When plotted in a semi-logarithmic scale,
the resulting $\phi$-dependence of $\bar{G}$ shows thus a characteristic double hump, commonly observed in 
granular metal films, which signals the metallic (percolating) and dielectric (tunneling) regimes.
The double hump feature, and the conductance step at $\phi\simeq \phi_c$, depend however on the ratio
$g_t/g_m$. For $g_t/g_m=1$, the EMA conductance decreases gradually as $\phi$ decreases without particular 
features at $\phi_c$, while a significant step becomes visible only for $g_t/g_m\ll 1$, as shown in Fig.~\ref{fig2}(b).

\subsection{Application to experiments}
\label{compa}
\begin{figure}[t]
\begin{center}
\includegraphics[scale=0.78,clip=true]{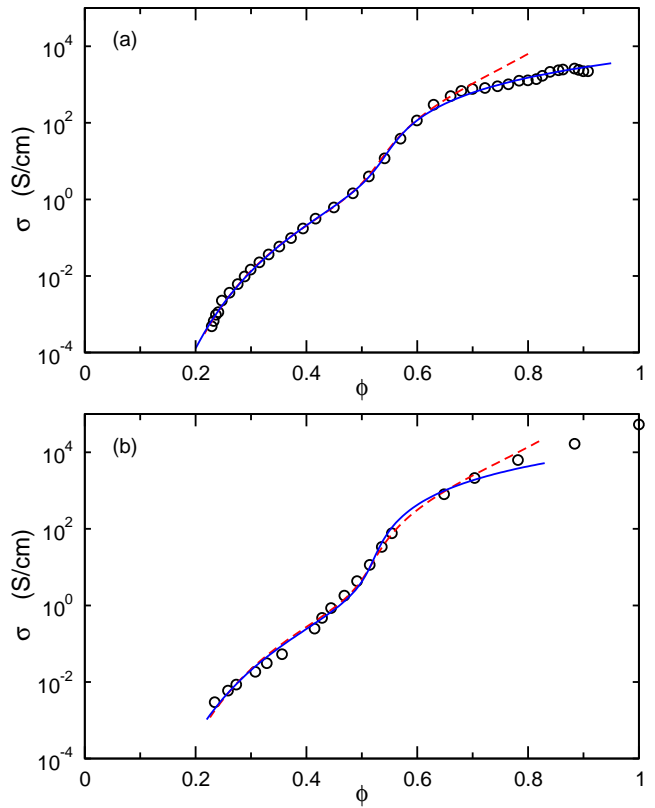}
\caption{(Color online) Measured conductivity $\sigma$ (open circles) as a function of Ni content for Ni-SiO$_2$
granular films. Data are taken (a) from Ref.~\onlinecite{Ni-SiO2} and (b) from Ref.~\onlinecite{Au-Al2O3}.
Solid lines are fitting curves from solutions of Eq.~\eqref{emamin}. Dashed lines are least-square fit results of the
cherry-pit EMA equation of Sec.~\ref{cherrypit}. Values of the fitting parameters are reported in Table~\ref{table1}.}
\label{fig3}
\end{center}
\end{figure}

\begin{figure}[t]
\begin{center}
\includegraphics[scale=0.78,clip=true]{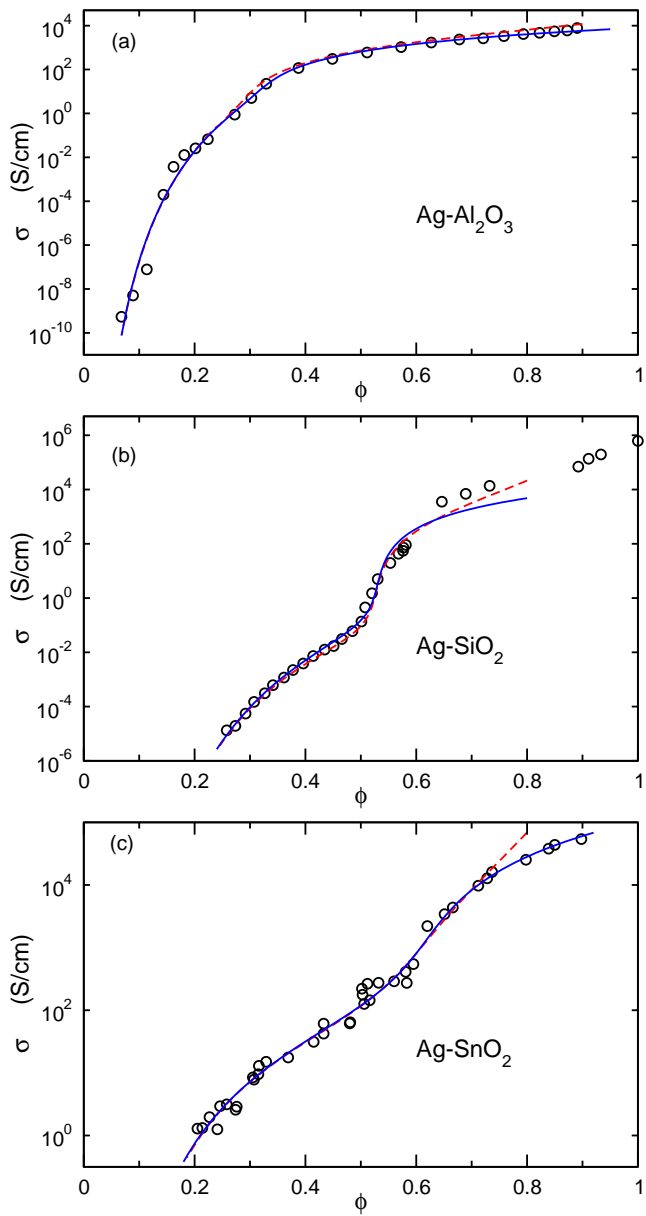}
\caption{(Color online) Measured conductivity $\sigma$ (open circles) as a function of Ag content for 
(a) Ag-Al$_2$O$_3$,\cite{Ag-Al2O3} (b) Ag-SiO$_2$,\cite{Ag/Au-SiO2} and (c) Ag-SnO$_2$,\cite{Ag-SnO2} 
granular films.  Solid lines are fitting curves from solutions of Eq.~\eqref{emamin}. Dashed lines are least-squares fit results of the
cherry-pit EMA equation of Sec.~\ref{cherrypit}. Values of the fitting parameters are reported in Table~\ref{table1}.}
\label{fig4}
\end{center}
\end{figure}

To assess the relevance of our EMA model for real nanocomposite films, we solve Eq.~\eqref{emamin}
so to reproduce published data of the conductivity of several granular metal systems. To this end,
we rewrite Eq.~\eqref{emamin} in terms of the dimensionless conductance $g^*=\bar{G}/g_m$ which,
besides $\phi$, depends on three parameters: $\phi_c$, $\xi/D$, and $g_t/g_m$. Since $g^*$ is independent
of the system size,\cite{Grimaldi2011,Wu2004} we define the EMA conductivity simply as $\bar{\sigma}=\Sigma g^*$, where
$\Sigma$ is a fourth fitting parameter which has the dimension of a conductivity. 

\begin{figure}[t]
\begin{center}
\includegraphics[scale=0.78,clip=true]{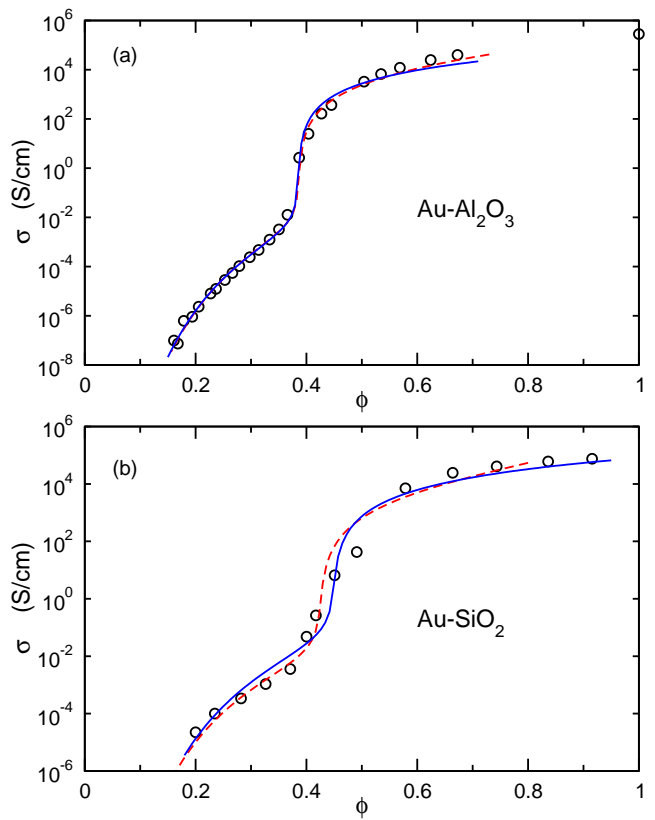}
\caption{(Color online) Measured conductivity $\sigma$ (open circles) as a function of Au content for 
(a) Au-Al$_2$O$_3$,\cite{Au-Al2O3} and (b) Au-SiO$_2$,\cite{Ag/Au-SiO2} granular films.  
Solid lines are fitting curves from solutions of Eq.~\eqref{emamin}. Dashed lines are least-squares fit results of the
cherry-pit EMA equation of Sec.~\ref{cherrypit}. Values of the fitting parameters are reported in Table~\ref{table1}.}
\label{fig5}
\end{center}
\end{figure}

To find the values of $\phi_c$, $\xi/D$, $g_t/g_m$, and $\Sigma$ which best fit the experimental data,
we apply a nonlinear least-squares algorithm to the numerical solution of Eq.~\eqref{emamin}. 
Results of this procedure applied to Ni-SiO$_2$ granular thick films are shown in Fig.~\ref{fig3}, where the 
EMA conductivity (solid lines) is fitted to the room temperature conductivity data of Ni-SiO$_2$ taken from 
Refs.~\onlinecite{Au-Al2O3,Ni-SiO2} (open circles). 
The fitted percolation threshold for the case of Fig.~\ref{fig3}(a), $\phi_c\simeq 0.52$, coincides 
with the value extracted in Ref.~\onlinecite{Ni-SiO2} from a fit with Eq.~\eqref{power}. 
This correspondence is not surprising because our EMA model has been constructed so as to reproduce the power-law 
behavior \eqref{power} with $t\simeq 2$ in the metallic regime. The value $\phi_c\simeq 0.51$ extracted
from the data of Ref.~\onlinecite{Au-Al2O3} indicates that the percolation threshold is independent of  
the conditions of the co-sputtering deposition, while these seem to affect to some extent the conductivity
above $\phi_c$.  

In the $\phi<\phi_c$ region, where tunneling dominates, the data from Ref.~\onlinecite{Ni-SiO2} 
are slightly better fitted than those from Ref.~\onlinecite{Au-Al2O3}. For Ni-SiO$_2$ and other 
granular metals, the mean size of the metallic particles decreases as $\phi$ is smaller,\cite{Au-Al2O3} while in 
our model we keep $D$ fixed. Hence, the better agreement of EMA for the case of Ref.~\onlinecite{Ni-SiO2}
could be attributed to a lower rate of decrease of $D$ than for the film of Ref.~\onlinecite{Au-Al2O3}.
We note that the resulting $\xi/D\simeq 0.045$ 
and $\xi/D\simeq 0.052$ extracted respectively from Figs.~\ref{fig3}(a) and \ref{fig3}(b) are nevertheless 
quite comparable, as also the tunneling to metal conductance ratio $g_t/g_m$ which is about $\sim 10^{-4}$
for both materials.

\begin{figure}[t]
\begin{center}
\includegraphics[scale=0.78,clip=true]{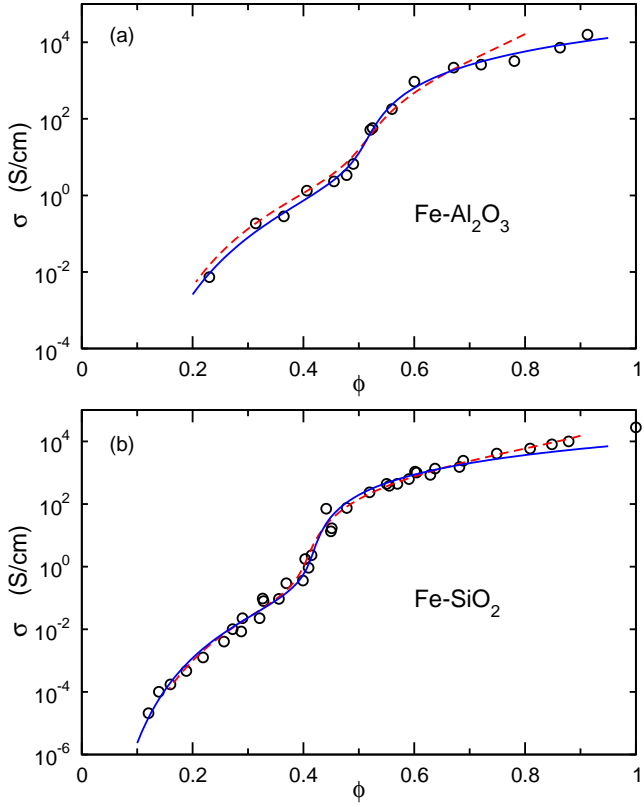}
\caption{(Color online) Measured conductivity $\sigma$ (open circles) as a function of Fe content for 
(a) Fe-Al$_2$O$_3$,\cite{Fe-Al2O3} and (b) Fe-SiO$_2$,\cite{Fe-SiO2} granular films.  
Solid lines are fitting curves from solutions of Eq.~\eqref{emamin}. Dashed lines are least-squares fit results of the
cherry-pit EMA equation of Sec.~\ref{cherrypit}. Values of the fitting parameters are reported in Table~\ref{table1}.}
\label{fig6}
\end{center}
\end{figure}

\begin{figure}[t]
\begin{center}
\includegraphics[scale=0.78,clip=true]{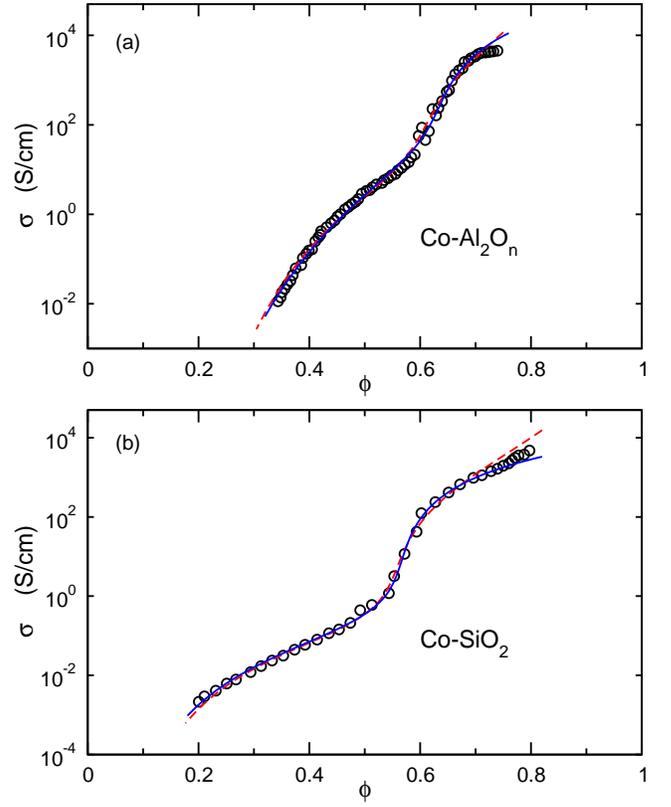}
\caption{(Color online) Measured conductivity $\sigma$ (open circles) as a function of Co content for 
(a) Co-Al$_2$O$_n$,\cite{Co-Al2On} and (b) Co-SiO$_2$,\cite{Co-SiO2} granular films.  
Solid lines are fitting curves from solutions of Eq.~\eqref{emamin}. Dashed lines are least-squares fit results of the
cherry-pit EMA equation of Sec.~\ref{cherrypit}. Values of the fitting parameters are reported in Table~\ref{table1}.}
\label{fig7}
\end{center}
\end{figure}

By following the same fitting procedure, we have reproduced the conductivity data of several
nanogranular films composed of noble metal\cite{Ag-SnO2,Ag-Al2O3,Ag/Au-SiO2,Au-Al2O3} or magnetic 
fillers\cite{Fe-Al2O3,Fe-SiO2,Co-Al2On,Co-SiO2} with different dielectric matrices, as shown 
in Figs.~\ref{fig4}-\ref{fig7}. The corresponding fitting parameters are reported in Table~\ref{table1}.
Despite the simplicity of Eq.~\eqref{emamin}, the overall quality of the fits is remarkable. In particular,
EMA captures well the dielectric regime below $\phi_c$ and the transition to the metallic regime
in the vicinity of $\phi_c$. 
Some deviations from the experimental data are visible in the large $\phi$ region, especially 
for Ag-SiO$_2$ in Fig.~\ref{fig4}(b), due to the imposed power-law behavior \eqref{power} which is expected 
to be valid only for $\phi$ immediately above $\phi_c$. 
Concerning the transition region, we point out that all films here considered 
have thicknesses in the micrometer range, justifying thus the use of the transport exponent value $t=2$ 
valid for three dimensional percolating systems. By using the EMA value $t=1$ for the transport exponent,
we obtain that the overall quality of the fits does not change appreciably: only in the transition region
about $\phi_c$ the fitting curves have occasionally a more abrupt variation, but the fitted values of $\phi_c$ 
and $\xi/D$ do not show appreciable variations.\cite{SupMat} 

From the values of $\phi_c$ reported in Table \ref{table1} we see that the percolation
threshold ranges between $0.3$ and $0.62$, with no correlation with the type of insulating phase. 
This result confirms earlier observations that the critical volume fraction depends on the particular 
combination of metal and insulator constituting the film.\cite{Au-Al2O3} Similarly, also the conductivity
step at about $\phi_c$, parametrized by $g_t/g_m$, does not show any particular trend. In this respect, 
we note that depending on the specific composite $g_t/g_m$ ranges between $\sim 10^{-3}$ and $\sim 10^{-7}$. As
mentioned previously, we expect $g_t$ to be smaller than $g_m$ due to particle interfacial barrier
and particle charging and Coulomb interaction effects. However, further reduction of $g_t$
can be induced also by non-random distributions of metal particle separations, as reported for example in Ref.~\onlinecite{Borziak1981} 
where gaps of the order of one nanometer in the interparticle spacing have been observed. For a nonzero
gap $\Delta$, indeed, we can replace the lowest limit of integration in the second term of Eq.~\eqref{ema1} by 
$D_\Delta=D+\Delta$.\cite{noterescale} Thus, if we rewrite the tunneling conductance in \eqref{modelg} as
\begin{equation}
\label{rescale}
g_t\exp\left[-\frac{2(r-D)}{\xi}\right]=g_t^*\exp\left[-\frac{2(r-D_\Delta)}{\xi}\right],
\end{equation}
where $g_t^*=\exp(-2\Delta/\xi)$ is a rescaled prefactor, the function $I$ of Eq.~\eqref{I} becomes:
\begin{equation}
\label{Irescaled}
I^*=8\phi\left(\frac{D_\Delta}{D}\right)^3\left\{\left[1+\frac{\xi}{2D_\Delta}\ln\!\left(\frac{g_t^*+\bar{G}}{\bar{G}}\right)\right]^3-1\right\},
\end{equation}
which in the dilute limit gives the same asymptotic $\bar{G}$ of Eq.~\eqref{expo3} with $g_t$ replaced by $g_t^*$.
If we interpret the values of $g_t/g_m$ reported in Table \ref{table1} as actually representing $g_t^*/g_m$, we
can easily explain values as small as $\sim 10^{-7}$, as observed for example for composites with Au. Indeed,
even assuming that $g_t\sim g_m$, from $g_t^*/g_m\sim \exp(-2\Delta/\xi)\sim 10^{-7}$ we get $\Delta\sim 0.8$ nm
for tunneling decay lengths of about $0.1$ nm.

\begin{table*}[t]
\caption{Values of $\phi_c$, $\xi/D$, $g_t/g_m$, and $\Sigma$ that best fit the measured conductivity data of 
Refs.~\onlinecite{Ni-SiO2,Ag-SnO2,Ag-Al2O3,Ag/Au-SiO2,Au-Al2O3,Fe-Al2O3,Fe-SiO2,Co-Al2On,Co-SiO2} obtained using the
EMA semi-phenomenological model of Sec.~\ref{minimal} (model A) and the EMA cherry-pit model of Sec.~\ref{cherrypit} (model B).
The values in parentheses are the fitted values of the impenetrability parameter $\lambda$ of the EMA cherry-pit model.}
\label{table1}
\begin{ruledtabular}
\begin{tabular}{lcccccccc}
Material & \multicolumn{2}{c}{$\phi_c$} & \multicolumn{2}{c}{$\xi/D$} & \multicolumn{2}{c}{$g_t/g_m$}
 & \multicolumn{2}{c}{$\Sigma$ (S/cm)}\\
         & model A & model B & model A & model B & model A & model B& model A & model B \\
\hline
Ni-SiO$_2$ (Ref.~\onlinecite{Ni-SiO2}) & $0.52$ & $0.49$ $(0.957)$ & $0.045$ & $0.046$ & $2.91\, 10^{-4}$ & $3.43\, 10^{-3}$ & $5.48\,10^{3}$ & $2.78\, 10^{2}$\\
Ni-SiO$_2$ (Ref.~\onlinecite{Au-Al2O3})& $0.51$ & $0.486$ ($0.954$) & $0.052$ & $0.051$ & $7.84\, 10^{-5}$ & $1.49\, 10^{-3}$ & $1.31\,10^{4}$ & $6.18\, 10^2$ \\
Ag-Al$_2$O$_3$ (Ref.~\onlinecite{Ag-Al2O3})& $0.3$ & $0.27$ ($0.58$) & $0.046$ & $0.047$ & $2.9\, 10^{-3}$ & $2.69\, 10^{-3}$ & $1.46\,10^{3}$ & $6.94\, 10^2$ \\
Ag-SiO$_2$ (Ref.~\onlinecite{Ag/Au-SiO2})& $0.52$ & $0.50$ ($0.96$)& $0.03$ & $0.032$& $6.26\, 10^{-6}$ & $4.73\, 10^{-5}$& $1.75\,10^{4}$ & $9.14\, 10^{2}$\\
Ag-SnO$_2$ (Ref.~\onlinecite{Ag-SnO2})& $0.59$ & $0.53$ ($0.97$)& $0.091$ & $0.094$ & $4.60\, 10^{-4}$ & $2.03\, 10^{-2}$ & $2.12\,10^{5}$ & $2.91\, 10^3$ \\
Au-Al$_2$O$_3$ (Ref.~\onlinecite{Au-Al2O3})& $0.38$ & $0.38$ ($0.862$)& $0.043$ & $0.045$ & $1.35\, 10^{-7}$ & $6.11\, 10^{-7}$ & $3.02\,10^{4}$ & $7.23\, 10^3$\\
Au-SiO$_2$ (Ref.~\onlinecite{Ag/Au-SiO2})& $0.45$ & $0.42$ ($0.90$)& $0.048$ & $0.056$& $7.74\, 10^{-7}$ & $2.45\, 10^{-6}$& $5.14\,10^{4}$ & $3.97\, 10^3$\\
Fe-Al$_2$O$_3$ (Ref.~\onlinecite{Fe-Al2O3})& $0.50$ & $0.48$ ($0.95$) & $0.061$ & $0.064$ & $1.29\, 10^{-4}$ & $3.14\, 10^{-3}$ & $1.60\,10^{4}$ & $7.74\, 10^2$ \\
Fe-SiO$_2$ (Ref.~\onlinecite{Fe-SiO2})& $0.41$ & $0.39$ ($0.878$)& $0.078$ & $0.078$& $2.89\, 10^{-5}$ & $2.04\, 10^{-4}$& $4.03\,10^{3}$ & $5.19\, 10^2$\\
Co-Al$_2$O$_n$ (Ref.~\onlinecite{Co-Al2On})& $0.62$ & $0.56$ ($0.98$)& $0.026$ & $0.026$& $1.22\, 10^{-4}$ & $3.96\, 10^{-3}$& $2.18\,10^{5}$ & $2\, 10^5$\\
Co-SiO$_2$ (Ref.~\onlinecite{Co-SiO2})& $0.56$ & $0.53$ ($0.97$)& $0.097$ & $0.095$& $1.02\, 10^{-5}$ & $2.7\, 10^{-4}$& $1.50\,10^{4}$ & $4.22\, 10^2$\\
\end{tabular}
\end{ruledtabular}
\end{table*}

Turning to the dielectric regime of $\sigma$ identified by the hump at $\phi<\phi_c$ in Figs.~\ref{fig3}-\ref{fig7}, 
we note that the tunneling decay length for a rectangular barrier is $\xi=\hbar/\sqrt{2m\varphi}$, 
where $m$ is the electron mass and $\varphi$ is the tunnel barrier height. We estimate $\varphi$ as the 
difference between the work function of the metal and the electron affinity of the dielectric. Since the work function for
the metals considered here ranges from about $4.5$ eV (Fe) to about $5.4$ eV (Au),\cite{Michaelson1977} 
while the electron affinities for Al$_2$O$_3$ and SiO$_2$ are respectively $\sim 1.35$ eV and $\sim 1$ eV,\cite{EA} 
we obtain that the tunneling decay length is $\xi\simeq 0.1$ nm for the composites with Al$_2$O$_3$ and SiO$_2$.
As the fitted values of $\xi/D$ range from $0.03$ to about $0.1$ (see Table~\ref{table1}), we infer that for this class 
of composites the mean size of metal particles is comprised
between $D \sim 1$ nm and $D \sim 3$ nm. This estimate is in fair accord with the observed particle sizes in 
these systems, as shown in Table~\ref{table2} where we compare our results of $D$ with measured values 
of the mean particle sizes.

For the Ag-SnO$_2$ system, the large electron affinity of the oxide semiconductor SnO$_2$ 
(about $4.3-4.5$ eV,\cite{elaffSnO2})  together with the work function $\sim 4.6$ eV for Ag,\cite{Michaelson1977}  
gives $\xi\simeq 0.4-0.7$ nm. From $\xi/D=0.09$ we obtain thus $D\simeq 4.4-7.8$ nm, which is comparable with 
$D\simeq 3-7$ nm measured in samples with $\phi<0.42$.\cite{Ag-SnO2} 
We note that using the point particle limit of Eq.~\eqref{power} (with $a_3\simeq 1.41$) to find $\xi/D$ from the conductivity 
data leads to the slightly larger estimate $\xi/D\simeq 0.13$.\cite{Ag-SnO2} 

\section{EMA cherry-pit model}
\label{cherrypit}

\begin{figure}[b]
\begin{center}
\includegraphics[scale=0.78,clip=true]{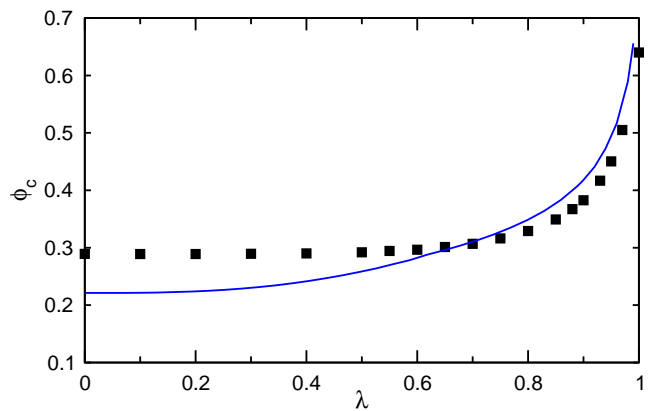}
\caption{(Color online) Critical volume fraction $\phi_c$ as a function of the impenetrability parameter 
$\lambda$ of the cherry-pit model. Solid line is the EMA $\phi_c$ obtained from $Z(\phi_c,D)=2$. Filled
squares are Monte Carlo results of Ref.~\onlinecite{Miller2009} in which Eqs.~\eqref{phi} and \eqref{A}
are used to relate $\phi_c$ with the critical density $\eta_c$.}
\label{fig8}
\end{center}
\end{figure}

The EMA model discussed in Sec.~\ref{minimal} treats the transition between the metallic and dielectric
regimes in a phenomenological way by introducing a critical volume fraction whose value is
found by fitting the experiments. In the model illustrated in Fig.~\ref{fig1} we have however assumed
that the metallic particles are allowed to overlap to some extent, and that the degree of overlapping
determines the value of $\phi_c$. Furthermore, in deriving Eq.~\eqref{emamin} we have considered the
metallic particles as completely uncorrelated by setting $g_2(r)=1$ for all particle contents lower than $\phi_c$.
To include explicitly particle overlaps and local correlation, we consider a cherry-pit model in which each metallic
sphere of diameter $D$ is composed by an impenetrable core of diameter $\lambda D$ surrounded by a penetrable
concentric shell of thickness $(1-\lambda)D/2$.\cite{TorquatoBook} Any two given metallic spheres can thus overlap as long as their
respective hard cores do not. The parameter $\lambda$ ranges between $0$ and $1$, which defines the limits of
fully penetrable and totally impenetrable spheres, respectively. For equilibrium distributions of cherry-pit 
spheres, the critical volume fraction for percolation of overlapping spheres varies thus between
$\phi_c\simeq 0.29$ for $\lambda=0$ and $\phi_c\simeq 0.64$ for $\lambda=1$,\cite{TorquatoBook} 
(see also Fig.~\ref{fig8}) consistently with the range of $\phi_c$ values we have obtained in Sec.~\ref{compa}.

\begin{table}[t]
\caption{Estimated particle sizes $D$ extracted from the $\xi/D$ values of Table~\ref{table1} using
$\xi=0.4-0.7$ nm for Ag-SnO$_2$ and $\xi=0.1$ nm for all other cases (see main text).
Unless otherwise indicated, the experimentally determined values of $D$ represent the mean particle 
sizes extracted from scanning or transmission electron microscopy in the dielectric regime 
(i.e., below the percolation threshold). Missing entries mean that the corresponding publications
do not report measurements of particle sizes for the granular film considered.}
\label{table2}
\begin{ruledtabular}
\begin{tabular}{lll}
Material & \multicolumn{2}{c}{$D$ (nm)} \\
         & Theory & Experiments    \\
         \hline
Ni-SiO$_2$ (Ref.~\onlinecite{Ni-SiO2}) & $2.2$  \\
Ni-SiO$_2$ (Ref.~\onlinecite{Au-Al2O3})& $2$ & $2.5\!-\!4$ (for $0.3\leq\phi\leq 0.55$)\\
Ag-Al$_2$O$_3$ (Ref.~\onlinecite{Ag-Al2O3})& $2.2$  \\
Ag-SiO$_2$ (Ref.~\onlinecite{Ag/Au-SiO2})& $3.3$ & $\sim 10$ (at $\phi=0.38$)\footnotemark[1]\\
Ag-SnO$_2$ (Ref.~\onlinecite{Ag-SnO2})& $4.4\!-\!7.8$ & $3\!-\!7$ (for $\phi< 0.42$)\\
Au-Al$_2$O$_3$ (Ref.~\onlinecite{Au-Al2O3})& $2.3$ & $1.8\!-\!3.5$ (for $0.15\leq\phi\leq 0.35$)\\
Au-SiO$_2$ (Ref.~\onlinecite{Ag/Au-SiO2})& $2$ & $2\!-\!8$(?) (for $0.1\leq\phi\leq 0.4$)\\
Fe-Al$_2$O$_3$ (Ref.~\onlinecite{Fe-Al2O3})& $1.7$ & $1\!-\!3$ (at $\phi=0.45$)\footnotemark[2]\\
Fe-SiO$_2$ (Ref.~\onlinecite{Fe-SiO2})& $1.3$ & $1\!-\!2.5$ (for $\phi\leq 0.3$)\\
Co-Al$_2$O$_n$ (Ref.~\onlinecite{Co-Al2On})& $3.8$ &  \\
Co-SiO$_2$ (Ref.~\onlinecite{Co-SiO2})& $1$ & $3.5\!-\!4.8$ (for $0.2\leq\phi\leq 0.35$)\footnotemark[2]\\
\end{tabular}
\end{ruledtabular}
\footnotetext[1]{Our estimate from Fig. 2(b) of Ref.~\onlinecite{Ag/Au-SiO2}.}
\footnotetext[2]{From fits of magnetization data.}
\end{table}

To apply the general EMA equation \eqref{ema1} to the case of equilibrium cherry-pit spheres, we note that
the radial distribution function $g_2(r)$ is that of hard-core spheres of diameter $\lambda D$, $g_2^\textrm{hc}(r;\lambda D)$,
as the penetrable shell has no effects on the equilibrium distribution.
Furthermore, to relate the fractional coverage $\phi$ of the cherry-pit spheres with the number density $\rho$, we use the
approximate but accurate formula:\cite{TorquatoBook,Kansal2002}
\begin{equation}
\label{phi}
\phi=1-(1-\eta\lambda^3)\exp\!\left[-\frac{(1-\lambda^3)\eta}{(1-\eta\lambda^3)^3}\right]A(\eta,\lambda),
\end{equation}
with
\begin{align}
\label{A}
A(\eta,\lambda)=&\exp\!\left\{-\frac{\eta^2\lambda^3(\lambda-1)}{2(1-\eta\lambda^3)^3}
[(7\lambda^2+7\lambda-2)\right.\nonumber \\
&-2\eta\lambda^3(7\lambda^2-5\lambda+1)+\eta^2\lambda^6(5\lambda^2-7\lambda+2)]\bigg\},
\end{align}
where we have introduced the dimensionless density $\eta=\pi D^3\rho/6$.\cite{noteeta} Equation \eqref{ema1} reduces in
this way to:
\begin{equation}
\label{emacp}
\frac{Z(\phi,D)g_m^{1/t}}{\bar{G}^{1/t}+g_m^{1/t}}
+\frac{24\eta}{D^3}\!\int_D^\infty\!dr r^2 \frac{g_2^\textrm{hc}(r;\lambda D)}{(\bar{G}/g_t)e^{\frac{2(r-D)}{\xi}}+1}=2,
\end{equation}
where we have corrected the first term by using the transport exponent $t$, as done in Eq.~\eqref{emamin}.
Since $g_2^\textrm{hc}(r;\lambda D)=0$ for $r<\lambda D$, the coordination number function for overlapping
spheres in Eq.~\eqref{emacp} is an integral between $\lambda D$ and $D$:
\begin{equation}
\label{zetacp}
Z(\phi,D)=\frac{24\eta}{D^3}\int_{\lambda D}^D\!dr r^2 g_2^\textrm{hc}(r;\lambda D).
\end{equation}
From the above equation we can already determine how the percolation threshold $\phi_c$ depends 
on the impenetrability parameter $\lambda$ by using the EMA relation $Z(\phi_c,D)=2$ derived
in Sec.~\ref{limiting}.
To this end, we use in Eq.~\eqref{zetacp} $g_2^\textrm{hs}(r;\lambda D)$  as
given by the accurate expression for the radial distribution function of hard spheres derived in 
Ref.~\onlinecite{Trokhy2005}, and apply Eqs.~\eqref{phi} and \eqref{A} to find $\phi_c$ from the critical
density $\eta_c$. The resulting critical volume fraction compares relatively well with the numerical calculations
for $\lambda\gtrsim 0.5$, as seen in Fig.~\ref{fig8} where the filled squares are the Monte Carlo results of
Ref.~\onlinecite{Miller2009}. 
By comparing the values of $\phi_c$ reported in Fig.~\ref{fig8} with those listed in Table~\ref{table1},
from which we see that $\phi_c>0.4$ with the exception of Ag-Al$_2$O$_3$, we infer 
that the percolation thresholds of nanogranular films are reproduced 
by the EMA cherry-pit model with $\lambda\gtrsim 0.85$,
which means that the spheres have generally little overlap. We obtain even (slightly) smaller overlaps if we compare
the experimental percolation thresholds with the Monte Carlo $\phi_c$ of Fig.~\ref{fig8}.

Although we do not expect that the detailed morphology of real granular metal films is fully reproduced
by equilibrium dispersions of cherry-pit spheres, these seem nevertheless to capture some critical aspects
of the microstructure and its evolution with $\phi$. We note also that other simple microscopic descriptions,
as for example the equilibrium permeable spheres model for which expressions of $g_2(r)$ and of the
volume fraction exist,\cite{permeable} may equally be used though they are of less practical 
implementation.

We proceed to apply the EMA cherry-pit model to the experimental data of Figs.~\ref{fig3}-\ref{fig7}
by using $\xi/D$, $g_t/g_m$, $\Sigma$, and $\lambda$ as fitting parameters and $t=2$ fixed.
We first invert numerically Eqs.~\eqref{phi} and \eqref{A} to extract $\eta$ from the measured $\phi$ values,
and subsequently we solve iteratively Eqs.~\eqref{emacp} and \eqref{zetacp} by using the model $g_2^\textrm{hs}(r;\lambda D)$
of Ref.~\onlinecite{Trokhy2005}. The results of nonlinear least-squares fits are shown by dashed lines
in Figs.~\ref{fig3}-\ref{fig7}, and the values of the fitting parameters that best reproduce the measured $\sigma$
are reported in Table \ref{table1}. We see that the cherry-pit model confirms the results obtained in
the previous section. In particular, the two fitting curves (solid and dashed lines) are practically
indistinguishable in both the dielectric regime and the transition region about $\phi_c$, and very similar values
of $\xi/D$ and $\phi_c$ are obtained from the two methods, as shown in Table~\ref{table1}. Furthermore, the 
quantitative accord with the experimental data for $\phi\lesssim\phi_c$ confirms
our assumption that local particle correlations are marginal in the dielectric regime. 
For metallic contents well above $\phi_c$, the conductivity of cherry-pit model is systematically larger
than that of the semi-phenomenological approach of Sec.~\ref{minimal} and gives occasionally better fits,
as seen in Figs.~\ref{fig3}(b), \ref{fig4}(b), and \ref{fig6}(b). We note however that the
radial distribution function of Ref.~\onlinecite{Trokhy2005} becomes inaccurate for packing fractions 
$\lambda^3\eta$ of the hard cores larger than $\sim 0.5$. Assuming for example $\lambda=0.95$, as is the case 
of Fe-Al$_2$O$_3$, in terms of the fractional coverage this limitation translates
into questionable results for $\phi\gtrsim 0.6$.

\section{Discussion and conclusions}
\label{concl}

The central result of this paper is that the two fundamentally different transport regimes of percolation 
and tunneling, which are simultaneously observed in many
conducting nanocomposite films, find a natural explanation within a single theoretical framework.
We have shown that quite general considerations on the nature of the interparticle electrical connectedness 
and on the distribution of the metallic phase are sufficient to describe quantitatively the dc conductivity $\sigma$
of several granular metal films. In particular, the semi-phenomenological
EMA equation derived in Sec.~\ref{minimal} represents a simple, yet efficient, tool to analyze the $\phi$-dependence of $\sigma$
and to estimate the values of the microscopic parameters that govern the observed conductivity behaviors.

In formulating the minimal model of Sec.~\ref{ema}, we have made different assumptions with the intent of keeping the theory 
as essential as possible. One of such assumptions concerns the direct tunneling decay of Eq.~\eqref{modelg} in which we
neglect particle charging and Coulomb interaction effects. As discussed in Sec.\ref{ema}, these become important as 
the temperature is lowered below room temperature and/or as the particle
size decreases.\cite{Beloborodov2007} For particle sizes of the order of a few nanometers, and well below the percolation
threshold, Coulomb effects become relevant 
also at room temperature, so that they could modify to some extent the $\xi/D$ values reported in Table \ref{table1}. 
A generalization of the present EMA approach as to include Coulomb gap effects would permit us to study on equal footing
both the concentration and the temperature dependencies of transport, while these two are generally treated 
separately. In this respect, measured $\sigma$ dependencies on both concentration and temperature, as those
reported for example in Refs.~\onlinecite{Au-Al2O3,Ni-SiO2,Ag-SnO2,Co-SiO2}, would find a more complete, and unified,
theoretical understanding.

In Sec.~\ref{ema} we have also assumed that the metallic particles are spherical and of fixed diameter. 
Although we do not expect that small deviations from sphericity would have any important effect,
metallic inclusions with high aspect-ratios can change appreciably the location of $\phi_c$ and
the low-density tunneling regime. For example, the tunneling conductivity of dispersions of rod particles of 
diameter $D$ and length $L\gg D$ scales approximately as $\sigma\propto\exp(-D^2/\xi\phi L)$ for isotropic 
orientation of rods.\cite{Ambrosetti2010a} The effect of elongated particles, as those observed in some granular
films with magnetic particles, can nevertheless be investigated by applying EMA to high aspect-ratio fillers,
as done for the tunneling case in Refs.~\onlinecite{Nigro2013b,Grimaldi2013}

Concerning the assumption of fixed particle size, we note that some composite films show a more or less 
pronounced reduction of the mean particle size $D$ as $\phi$ decreases,
as reported for example in Ref.~\onlinecite{Au-Al2O3} and in Table~\ref{table2}. This effect can be included in 
Eq.~\eqref{ema1} by considering an explicit $\phi$-dependence of $D$ which simulates the observed one. In principle, it is possible
to consider within EMA also the effect of particle size polydispersity, although this would require detailed 
knowledge of the size distribution and its possible dependence on $\phi$.\cite{Labarta2006} In the absence of
these informations, the theoretical estimates of Table~\ref{table2} can be tentatively interpreted in terms of effective sizes $D_\textrm{eff}$ of polydisperse particles. It is not difficult to estimate $D_\textrm{eff}$ from the solution of 
Eq.~\eqref{expo2} for asymptotically small $\bar{G}$, which is given by the last line of Eq.~\eqref{expo3} with
$D$ replaced by $D_\textrm{eff}=\sqrt[3]{\langle D^3\rangle}$. Particle size distributions with long tails for large $D$
may thus have $D_\textrm{eff}$ considerably larger than the mean $\langle D\rangle$.
We note however that to coherently describe the effect of particle size polydispersity, charging and Coulomb 
interactions should be considered as well, since these become increasingly important as particle sizes are smaller.

Finally, we point out that although the cherry-pit model of Sec.~\ref{cherrypit} includes local correlations induced by
the particle hard-cores, it ignores possible long-range correlations and is not suitable to describe particle clustering 
or aggregation effects.
Although the granular films here considered do not appear to show long-range correlations, the general two-site EMA equation
\eqref{ema1} allows us to include at least partially these effects through suitable choices of the radial distribution function $g_2(r)$. 
Aggregation induced by effective particle attractions can be modeled for example by attractive square-well potentials, for which
approximate expressions for $g_2(r)$ are available.\cite{Yuste1994} In the case of tunneling, the resulting EMA conductance is in
excellent overall agreement with numerical simulations for a wide range of potential profiles, as shown in 
Ref.~\onlinecite{Nigro2012b}. Another possible route to simulate phenomenologically particle aggregation and 
clustering is to consider simple square-well models of $g_2(r)$,\cite{Nigro2012b,Grimaldi2013} whose values for $r$
lower and larger than some characteristic correlation distance can be adjusted to fit the measured conductivity.

We conclude by mentioning that even though we have formulated the theory to describe granular thick films
as three dimensional systems, it is in principle not difficult to modify EMA to (quasi) two-dimensional systems,
so to describe transport in thin granular films as those studied for example in Refs.~\onlinecite{Salvadori2013,Fostner2014}. 

\acknowledgements

The author thanks I. Balberg, A. P. Chatterjee, and M. C. Salvadori for useful 
comments and Z.-Q. Li for kindly providing the original conductivity data of Ref.~\onlinecite{Ag-SnO2}.

\widetext

\section{Supplemental Material for ``Theory of percolation and tunneling regimes in nanogranular metal films''}
To evaluate the effect of varying the transport exponent $t$ in the EMA model of Sec.~III,
we fit the experimental conductivity data of 
Refs.~\onlinecite{Ni-SiO2,Ag-SnO2,Ag-Al2O3,Ag/Au-SiO2,Au-Al2O3,Fe-Al2O3,Fe-SiO2,Co-Al2On,Co-SiO2}
with the EMA conductivity resulting from the solution of
\begin{equation}
\label{emamin}
\frac{(\phi/\phi_c)g_m^{1/t}}{\bar{G}^{1/t}+g_m^{1/t}}+
4\phi\left\{\left[1+\frac{\xi}{2D}\ln\!\left(\frac{g_t+\bar{G}}{\bar{G}}\right)\right]^3-1\right\}=1,
\end{equation}
with transport exponent fixed at $t=2$ and $t=1$. The case $t=2$ corresponds to the approach followed in Sec.~III in which
$t$ is adjusted to reproduce the percolation exponent for three dimensional systems. The case $t=1$ is the EMA
exponent. The resulting fits are shown in Figs.~\ref{fig3}-\ref{fig7},
and the corresponding values of $\phi_c$, $\xi/D$, $g_t/g_m$, and $\Sigma$ are reported in Table~\ref{table1}.

\begin{figure}[h]
\begin{center}
\includegraphics[scale=0.68,clip=true]{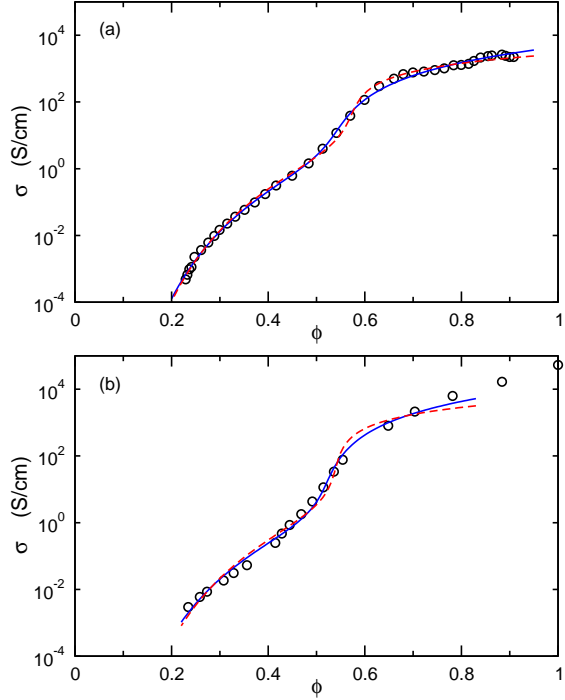}
\caption{(Color online) Measured conductivity $\sigma$ (open circles) as a function of Ni content for Ni-SiO$_2$
granular films. Data are taken (a) from Ref.~\onlinecite{Ni-SiO2} and (b) from Ref.~\onlinecite{Au-Al2O3}.
Solid (dashed) lines are fitting curves from solutions of Eq.~\eqref{emamin} with $t=2$ ($t=1$). }
\label{fig3}
\end{center}
\end{figure}

\begin{figure}[t]
\begin{center}
\includegraphics[scale=0.68,clip=true]{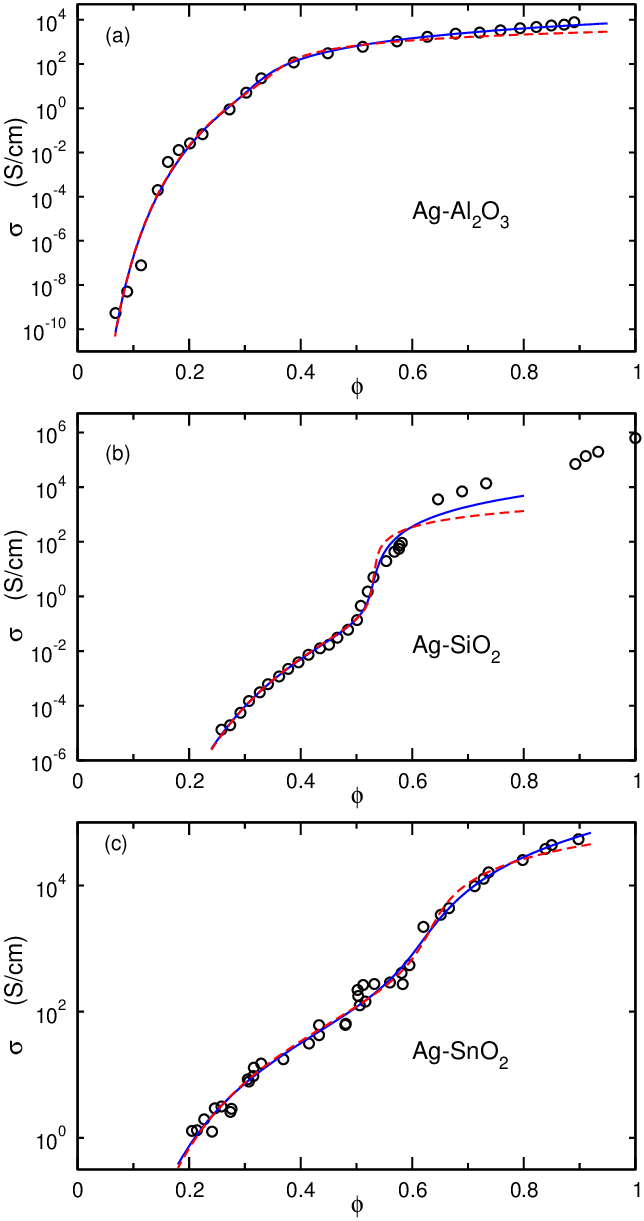}
\caption{(Color online) Measured conductivity $\sigma$ (open circles) as a function of Ag content for 
(a) Ag-Al$_2$O$_3$,\cite{Ag-Al2O3} (b) Ag-SiO$_2$,\cite{Ag/Au-SiO2} and (c) Ag-SnO$_2$,\cite{Ag-SnO2} 
granular films.  Solid (dashed) lines are fitting curves from solutions of Eq.~\eqref{emamin} with $t=2$ ($t=1$). }
\label{fig4}
\end{center}
\end{figure}
\begin{figure}[t]
\begin{center}
\includegraphics[scale=0.68,clip=true]{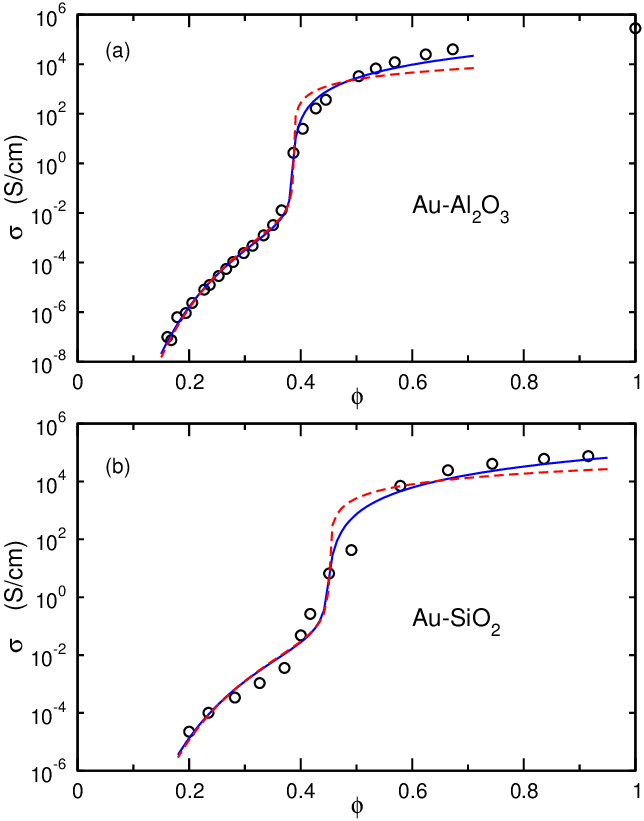}
\caption{(Color online) Measured conductivity $\sigma$ (open circles) as a function of Au content for 
(a) Au-Al$_2$O$_3$,\cite{Au-Al2O3} and (b) Au-SiO$_2$,\cite{Ag/Au-SiO2} granular films.  
Solid (dashed) lines are fitting curves from solutions of Eq.~\eqref{emamin} with $t=2$ ($t=1$). }
\label{fig5}
\end{center}
\end{figure}
\begin{figure}[t]
\begin{center}
\includegraphics[scale=0.68,clip=true]{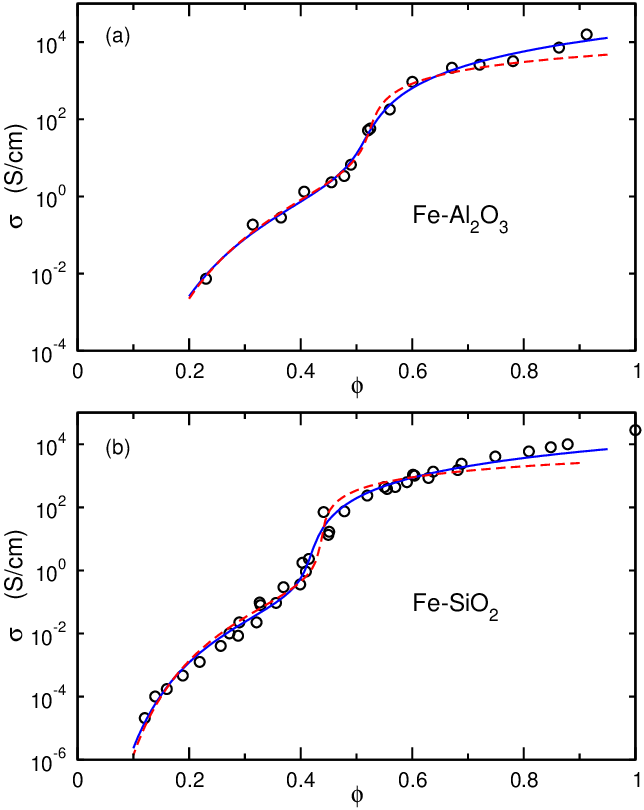}
\caption{(Color online) Measured conductivity $\sigma$ (open circles) as a function of Fe content for 
(a) Fe-Al$_2$O$_3$,\cite{Fe-Al2O3} and (b) Fe-SiO$_2$,\cite{Fe-SiO2} granular films.  
Solid (dashed) lines are fitting curves from solutions of Eq.~\eqref{emamin} with $t=2$ ($t=1$). }
\label{fig6}
\end{center}
\end{figure}
\begin{figure}[t]
\begin{center}
\includegraphics[scale=0.68,clip=true]{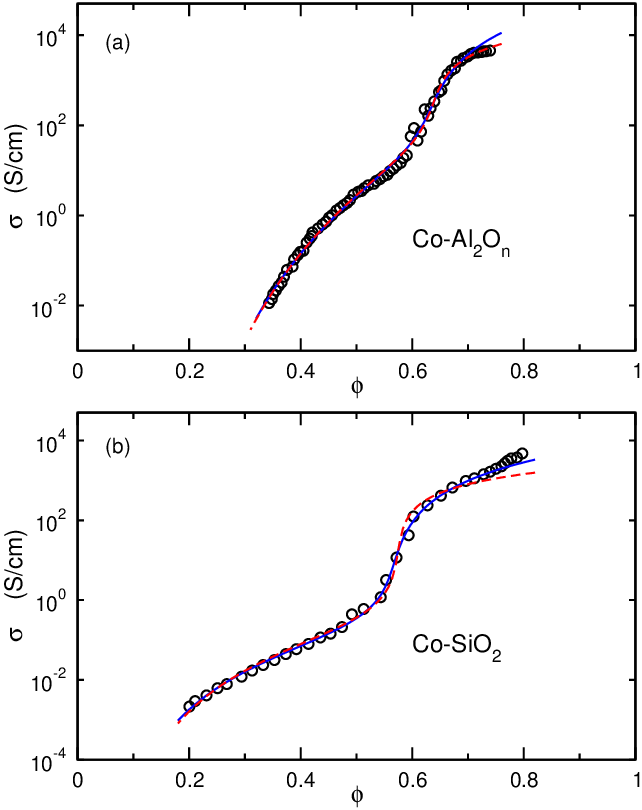}
\caption{(Color online) Measured conductivity $\sigma$ (open circles) as a function of Co content for 
(a) Co-Al$_2$O$_n$,\cite{Co-Al2On} and (b) Co-SiO$_2$,\cite{Co-SiO2} granular films.  
Solid (dashed) lines are fitting curves from solutions of Eq.~\eqref{emamin} with $t=2$ ($t=1$). }
\label{fig7}
\end{center}
\end{figure}

\begin{table*}[t]
\caption{Values of $\phi_c$, $\xi/D$, $g_t/g_m$, and $\Sigma$ that best fit the measured conductivity data of 
Refs.~\onlinecite{Ni-SiO2,Ag-SnO2,Ag-Al2O3,Ag/Au-SiO2,Au-Al2O3,Fe-Al2O3,Fe-SiO2,Co-Al2On,Co-SiO2} obtained using the
EMA semi-phenomenological model of Eq.~\eqref{emamin} with $t=2$ and $t=1$.}
\label{table1}
\begin{ruledtabular}
\begin{tabular}{lcccccccc}
Material & \multicolumn{2}{c}{$\phi_c$} & \multicolumn{2}{c}{$\xi/D$} & \multicolumn{2}{c}{$g_t/g_m$}
 & \multicolumn{2}{c}{$\Sigma$ (S/cm)}\\
         & $t=2$ & $t=1$ & $t=2$ & $t=1$ & $t=2$ & $t=1$ & $t=2$ & $t=1$ \\
\hline
Ni-SiO$_2$ (Ref.~\onlinecite{Ni-SiO2}) & $0.52$ & $0.57$ & $0.045$ & $0.042$ & $2.91\, 10^{-4}$ & $1.10\, 10^{-3}$ & $5.48\,10^{3}$ & $3.63\, 10^{3}$\\
Ni-SiO$_2$ (Ref.~\onlinecite{Au-Al2O3})& $0.51$ & $0.54$ & $0.052$ & $0.047$ & $7.84\, 10^{-5}$ & $4.07\, 10^{-4}$ & $1.31\,10^{4}$ & $5.92\, 10^3$ \\
Ag-Al$_2$O$_3$ (Ref.~\onlinecite{Ag-Al2O3})& $0.3$ & $0.36$ & $0.046$ & $0.044$ & $2.9\, 10^{-3}$ & $1.84\, 10^{-2}$ & $1.46\,10^{3}$ & $1.78\, 10^3$ \\
Ag-SiO$_2$ (Ref.~\onlinecite{Ag/Au-SiO2})& $0.52$ & $0.53$ & $0.03$ & $0.03$ & $6.26\, 10^{-6}$ & $5.32\, 10^{-5}$& $1.75\,10^{4}$ & $2.63\, 10^{3}$\\
Ag-SnO$_2$ (Ref.~\onlinecite{Ag-SnO2})& $0.59$ & $0.64$ & $0.091$ & $0.085$ & $4.60\, 10^{-4}$ & $1.52\, 10^{-3}$ & $2.12\,10^{5}$ & $1\, 10^5$ \\
Au-Al$_2$O$_3$ (Ref.~\onlinecite{Au-Al2O3})& $0.38$ & $0.39$ & $0.043$ & $0.041$ & $1.35\, 10^{-7}$ & $7.12\, 10^{-7}$ & $3.02\,10^{4}$ & $8.36\, 10^3$\\
Au-SiO$_2$ (Ref.~\onlinecite{Ag/Au-SiO2})& $0.45$ & $0.45$ & $0.048$ & $0.046$& $7.74\, 10^{-7}$ & $2.13\, 10^{-6}$& $5.14\,10^{4}$ & $2.43\, 10^4$\\
Fe-Al$_2$O$_3$ (Ref.~\onlinecite{Fe-Al2O3})& $0.50$ & $0.53$ & $0.061$ & $0.057$ & $1.29\, 10^{-4}$ & $5.8\, 10^{-4}$ & $1.60\,10^{4}$ & $5.85\, 10^3$ \\
Fe-SiO$_2$ (Ref.~\onlinecite{Fe-SiO2})& $0.41$ & $0.44$ & $0.078$ & $0.071$ & $2.89\, 10^{-5}$ & $1.2\, 10^{-4}$& $4.03\,10^{3}$ & $2.41\, 10^3$\\
Co-Al$_2$O$_n$ (Ref.~\onlinecite{Co-Al2On})& $0.62$ & $0.64$ & $0.026$ & $0.026$ & $1.22\, 10^{-4}$ & $1.2\, 10^{-3}$& $2.18\,10^{5}$ & $3.34\, 10^4$\\
Co-SiO$_2$ (Ref.~\onlinecite{Co-SiO2})& $0.56$ & $0.57$ & $0.097$ & $0.089$& $1.02\, 10^{-5}$ & $6.1\, 10^{-5}$& $1.50\,10^{4}$ & $3.66\, 10^3$\\
\end{tabular}
\end{ruledtabular}
\end{table*}


\begin{thebibliography}{50}
\bibitem{noterescale}
This argument is equivalent to considering a gapped radial distribution function such that
$g_2(r)=0$ for $D\leq r\leq D_\Delta$ and $g_2(r)\neq 0$ for $r>D_\Delta$.
\bibitem{noteeta}
From Eqs.~\eqref{phi} and \eqref{A} we recover the fully penetrable sphere limit
$\phi=1-\exp(-\eta)$ for $\lambda=0$ and the fully impenetrable (hard-core) sphere
limit $\phi=\eta$ for $\lambda=1$.
\end{thebibliography}
\end{document}